\documentclass[twocolumn]{article}
\usepackage{graphicx}
\makeatletter
\def\@normalsize{\@setsize\normalsize{10pt}\xpt\@xpt
\abovedisplayskip 10pt plus2pt minus5pt\belowdisplayskip \abovedisplayskip
\abovedisplayshortskip \z@ plus3pt\belowdisplayshortskip 6pt plus3pt
minus3pt\let\@listi\@listI}
\def\subsize{\@setsize\subsize{12pt}\xipt\@xipt}
\def\section{\@startsection {section}{1}{\z@}{1.0ex plus 1ex minus
 .2ex}{.2ex plus .2ex}{\large\bf}}
\def\subsection{\@startsection {subsection}{2}{\z@}{.2ex plus 1ex}
{.2ex plus .2ex}{\subsize\bf}}
\makeatother
\begin{document}
%don't want date printed
\date{}
%make title bold and 14 pt font (Latex default is non-bold, 16pt)
\title{\Large\bf Maxwell $F^N$ Characteristic Equation Algorithm Applied to Abelian Born-Infeld Action in D$p$-branes}
%for single author (just remove % characters)
\author{Joseph Ambrose G. Pagaran\\
  Department of Physics\\
  School of Science and Engineering\\
  Ateneo de Manila University\\
  1108 Loyola Heights, Quezon City}
%for two authors (this is what is printed)
%\author{\begin{tabular}[t]{c@{\extracolsep{8em}}c}
%I. M. Author  & M. Y. Coauthor \\
% \\
%        My Department & Coauthor Department \\
%        My Institute & Coauthor Institute \\
%        author@domain.name & coauthor@domain.name
%\end{tabular}}
\maketitle
%I don't know why I have to reset thispagestyle, but otherwise get page numbers
%\thispagestyle{empty}
\subsection*{Abstract}
%IEEE allows italicized abstract
{\em An algorithm is devised to generate characteristic identities
between Maxwell fieldstrength invariants (traced over Lorentz
indices and disregarding ordering) that suffer linear dependence
in certain dimensionalities as they have been originally obtained
using a Maple routine. These relations between invariants are then
applied to simplify the Abelian Born-Infeld (ABI) effective action
in arbitrary degree of fieldstrength invariants. I have explicitly
displayed the simplified ABI action in 4, 6, 8, 10, and 12
space-time dimensions relevant in D$p$-branes.
%end italics mode
}

\section{Introduction}
Abelian Born-Infeld (ABI) action \cite{born} is the non-linear
generalization of the Maxwell action in quantum electrodynamics.
It appears as a low energy effective action of open strings and in
world-volume effective action in $D$-branes \cite{ket}.

In particular, a flat $D$-brane in type II string theory the
bosonic massless degrees of freedom of an open string ending on
the $D$-brane are a U(1) gauge field. Apart from the 9-$p$ neutral
scalar fields that describe the transversal excitations of the
$D$-branes which I choose to disregard, this U(1) gauge field in
$p+1$ dimensions describes the open string longitudinal
fluctuations of the brane.\cite{pol} When $n$ $D$-branes coincide,
the massless degrees of freedom of open strings beginning and
ending on them are a U($n$) gauge field. \cite{wit} (A number of
the disregarded scalar fields are in the adjoint of the gauge
group.)

The present work explores the structure of the ABI action
\cite{sev}
\begin{equation}\label{abi}
    \Gamma_{\mathrm{BI}}
    =-T_p\int d^{p+1}\!\sigma\,\,
    \sqrt{
        \det(
            \delta_{{\mu}{\nu}}-2\pi\alpha'F_{{\mu}{\nu}})}
\end{equation}
up to all orders in the string length $\sqrt{\alpha'}$ in the
limit of slowly varying fieldstrengths, in the assumption of zero
derivatives of the fieldstrength. Furthermore, the static gauge is
choosen and transverse scalars are disregarded throughout the
paper. Only dynamics contributed by the longitudinal excitations
are considered as they are described by the U($n$) gauge of $n$
several coinciding $D$-branes. Here in (\ref{abi}), $T_p$ is the
D$p$-brane tension, $\alpha\epsilon\{0,1,\ldots,p\}$.

Now in the ABI (\ref{abi}) action, the term can be expanded in
powers of $F$ \cite{tsey}
\begin{equation}\label{sqrt}
    \sqrt{
        \det(
            \delta_{{\mu}{\nu}}+F_{{\mu}{\nu}})}
    \!=\!\sum^\infty_{k=0}
    C^{\mu_1\nu_1\ldots\mu_k\nu_k}F_{\mu_1\nu_1}\cdot\cdot\cdot F_{\mu_{2k}\nu_{2k}}
\end{equation}
 where the Lorentz tensors
$C^{\mu_1\nu_1\ldots\mu_k\nu_k}$  are defined as coefficients in
the $2\pi\alpha'=1$ setting. Here, such coefficients are set to 1
directing then our attention to the fieldstrength tensor in powers
of $F$, $F^{2k}$, and investigate its structure.

These powers of $F$ in (\ref{sqrt}) has been given attention in
the work of Delbourgo \cite{delb} where characteristic equations
were expressed in terms of polynomials over traces of matrix
powers. In this work, he developed a Maple routine that computes a
set of invariants involving the electromagnetic Maxwell field
tensor in arbitrary space-time dimensions specifically for
describing multiphoton processes \cite{ritz}.

Here, I devised an algorithm that generates the same set of
invariants without using any symbolic software. But I use the
results obtained from the latter and constructed a prescription
based on the principle of induction. The devised algorithm will be
enumerated in Section 2. A test run of the algorithm is performed
in Section 3 at the same time compact notations are introduced.
After having explicitly displayed the characteristic equations,
the same set of equations is used to simplify the powers of $F$ in
the ABI action as it has been expanded in Eqn (\ref{sqrt}). I
displayed explicitly in Section 4 the simplified version of the
ABI action relevant to D$p$-branes in 4, 6, 8, 10, and 12
space-time dimensions.

\section{The Algorithm}
The following algorithm is devised to construct characteristic
equations relating fieldstrength invariants
\begin{equation}\label{fn}
    \sum^{d=\mathrm{max}(2e+f)}_{e=0}T_{(2e)}F^f=0
\end{equation}
($T_{(0)}=1$) that suffer linear dependence in certain
dimensionalities. That is,
\begin{eqnarray}\label{fv}
    F^d\!\!\!\!\!&=&\!\!\!\!\!T_{(2)}F^{d-2}+T_{(4)}F^{d-4}+\ldots+T_{(d)}\\
    \label{fd}
    F^{d}\!\!\!\!\!&=&\!\!\!\!\!T_{(2)}F^{d-2}+T_{(4)}F^{d-4}+\ldots+T_{(d-1)}F
\end{eqnarray}
when $d$ is even (\ref{fv}) and odd (\ref{fd}), respectively.
These are subject to the condition that the degree $d$ of $F$ be
$d=\mbox{max}(2e+f)$. Furthermore, $T_{(2e)}$ are the terms the
devised algorithm generate.
\begin{enumerate}
\item[(a)] All of the terms within a characteristic equation are all of degree
$d$. For example:
\begin{equation}
    F^{12},\,\,\,T_2T_4F^6,\,\,\,T_4F^8,\,\,\,T_2T_4T_6,\,\,\,\mbox{etc.}
\end{equation}
are all of degree 12.
\item[(b)] Each characteristic equation has one fieldstrength tensor of
degree $D$, say for $D=d$, the relevant tensor is $F^d$. See
example in (a).
\item[(c)]  With $T_{j}=\mbox{Tr}(F^j)$, the rest
of the terms in each characteristic equation are products of the
form $\prod_kT_{j_k}$ where $j=j_k$ is even only and that $\sum_k
j_k =d$ is also even. The subscript here initially is set to
$j=d$. See example in (a).
\item[(d)] These are only formed when the
degree is even also. It is not constructed when the degree is odd.
The following are all of degree 13,
\begin{equation}
    F^{13},\,\,\,T_2T_4F^7,\,\,\,T_4F^9,\,\,\,T_2T_4T_6F,\,\,\,\mbox{etc.}
\end{equation}
which is Example in (a) times $F$.
\item[(e)] Each term formed in steps (c)-(d) has a prefactor
$(-1)^t\frac{1}{1!j^1}$ with $j$ and $d$ defined in (c). This is
applied only when $j_k$ in (c) is unique. Here $t$ denotes the
degree of $T$.
\item[(f)] If in $\prod_kT_{j_k}$, $j_{k}$ is not unique (i.e.
$j_{k}$ appeared as a subscript $p$ times) the term formed in step
(e) will have a prefactor $(-1)^t\frac{1}{p!j^p}$. It can be a
product $(-1)^d\prod_{q}\frac{1}{(p_q)!j^{p_q}}$ for every $q$th
$T_j$ term that appeared $p_q$ times. It should be noted that
$T_2T_2T_2$ will have a different prefactor with $T^3_2$. And that
the latter expression should be used.
\item[(g)]
Characteristic equations whose common degree is odd is simply the
even degree characteristic equation (obtained from steps (a)-(c))
with all terms unit $F$ multiplied.
\item[(h)] All terms in a characteristic equation whose degree $d+1$ is odd
are all carried to $d+2$ characteristic equation except that $F$
is multiplied to each term. The additional degree is contributed
by this unit $F$ multiplication. Since $d+2$ is again even, steps
(c)-(f) is repeated.
\item[(i)] All terms formed in the above steps are added after following
strictly the constraints provided in each step as in (\ref{fn}) or
equivalently (\ref{fv}) or (\ref{fd}). These terms comprise the
characteristic equation of $F^n$ true only when $F$ is abelian.
\end{enumerate}
It should be noted that this algorithm is constructed in the
assumption of zero derivatives of the fieldstrength tensor and no
ordering in indices is set. No other constraints about the indices
are imposed.

\section{Test Run of Algorithm}
Applying the algorithm in the Section 2 (particularly steps (c) to
(f)), unique traced invariants $T_{(d)}$ for even $d$ (where
$d=2,4,6,8,10$) are displayed as follows.
\begin{eqnarray}
%  \begin{array}{l}
    &&\!\!\!\!\!\!\!\!\!\!\!\!\!\!\!\!\!\!\!\!\!\!
        T_{(2)}\!=\!+\frac{1}{2}T_2,%\nonumber\\
    %&&\!\!\!\!\!\!\!\!\!\!\!\!\!\!\!\!\!\!\!\!\!\!
    \,\,\,
        T_{(4)}\!=\!-\frac{1}{8}T^2_2+\frac{1}{4}T_4\nonumber\\
    &&\!\!\!\!\!\!\!\!\!\!\!\!\!\!\!\!\!\!\!\!\!\!
        T_{(6)}\!=\!
            +\frac{1}{48}T^3_2
            -\frac{1}{8}T^{}_{2}T^{}_{4}+\frac{1}{6}T_6\label{t}\\\nonumber
    &&\!\!\!\!\!\!\!\!\!\!\!\!\!\!\!\!\!\!\!\!\!\!
        T_{(8)}\!=\!
            -\frac{1}{384}T^4_2
            +\frac{1}{32}T^{2}_{2}T^{}_{4}
        -\frac{1}{12}T^{}_{2}T^{}_{6}
        -\frac{1}{32}T_4^2+\frac{1}{8}T_8\\
    &&\!\!\!\!\!\!\!\!\!\!\!\!\!\!\!\!\!\!\!\!\!\!
    T_{(10)}\!=\!+\frac{1}{3840}T^5_2
            -\frac{1}{192}T^{3}_{2}T^{}_{4}
            +\frac{1}{48}T^{2}_{2}T^{}_{6}
            -\frac{1}{16}T^{}_{2}T^{}_{8}
    \nonumber\\&&
    \,\,\,\,\,\,\,\,\,\,\,\,\,\,\,\,\,\,\,\,\,\,\,\,\,\,\,\,\,\,\,\,\,\,\,\,\,\,\,\,\,\,\,\,\,\,\,\,\,\,\,\,\,\,\,\,\,\,\,\,\,\,
            -\frac{1}{24}T^{}_{4}T^{}_{6}
    +\frac{1}{10}T_{10}\nonumber
%  \end{array}
\end{eqnarray}
where $\sum_k a_k\cdot b_k=d$ for even $d$ in $T^{a_k}_{b_k}$.
While $d=2,4,6$ is a reproduction of the results obtained via a
Maple routine \cite{delb}, the same characteristic invariants
(\ref{fn}) were obtained following the algorithm devised in
Section 2. Here, $T_{(2e)}$ are given in Eqn (\ref{t}). Explicitly
(\ref{fn}) (or equivalently (\ref{fv}) or (\ref{fd})), these are
($d=2$ up to $d=10$)
\begin{eqnarray}\label{fnexp}
%  \begin{array}{l}
    &&\!\!\!\!\!\!\!\!\!\!\!\!\!\!\!\!\!\!\!\!\!\!
    F^2\!-\!T_{(2)}\!\!=\!0, \nonumber\\
    &&\!\!\!\!\!\!\!\!\!\!\!\!\!\!\!\!\!\!\!\!\!\!
    \,
    F^3\!-\!T_{(2)}F\!\!=\!0,\nonumber\\
    &&\!\!\!\!\!\!\!\!\!\!\!\!\!\!\!\!\!\!\!\!\!\!
    \,
    F^4\!-\!T_{(2)}F^2\!-\!T_{(4)}\!\!=\!0\nonumber\\
    &&\!\!\!\!\!\!\!\!\!\!\!\!\!\!\!\!\!\!\!\!\!\!
    F^5\!-\!T_{(2)}F^3\!-\!T_{(4)}F\!\!=\!0,\nonumber\\
    &&\!\!\!\!\!\!\!\!\!\!\!\!\!\!\!\!\!\!\!\!\!\!
    F^6\!-\!T_{(2)}F^4\!-\!T_{(4)}F^2\!-\!T_{(6)}\!\!\!=\!0\nonumber\\
    &&\!\!\!\!\!\!\!\!\!\!\!\!\!\!\!\!\!\!\!\!\!\!
    F^7\!-\!T_{(2)}F^5\!-\!T_{(4)}F^3\!-\!T_{(6)}F\!\!\!=\!0\\
    &&\!\!\!\!\!\!\!\!\!\!\!\!\!\!\!\!\!\!\!\!\!\!
    F^8\!-\!T_{(2)}F^6\!-\!T_{(4)}F^4\!-\!T_{(6)}F^2\!-\!T_{(8)}\!\!\!=\!0\nonumber\\
    &&\!\!\!\!\!\!\!\!\!\!\!\!\!\!\!\!\!\!\!\!\!\!
    F^9\!-\!T_{(2)}F^7\!-\!T_{(4)}F^5\!-\!T_{(6)}F^3\!-\!T_{(8)}F\!=\!0\nonumber\\
    &&\!\!\!\!\!\!\!\!\!\!\!\!\!\!\!\!\!\!\!\!\!\!
    F^{10}\!-\!T_{(2)}F^8\!-\!T_{(4)}F^6\!-\!T_{(6)}F^4\!-\!T_{(8)}F^2\!-\!T_{(10)}\!=\!0\nonumber%\\\nonumber
%  \end{array}
\end{eqnarray}
The case when $d=8,9,10$ is shown explicitly for illustration
purposes. In Eqn (\ref{fnexp}), $F^n$ in odd degrees are now
displayed illustrating the application of steps (d), (g) and (i).
These correspond to $D=2,4,6,8,10,12,14,16,18,20$ space-time
dimensions, respectively. For my purposes, I consider $D=2$ up to
$D=12$ space-time dimensions.

\section{Application to Abelian Born-Infeld Action}
Now that I have the algorithm in place, let me consider the
following notations, conventions, and definitions as well as
assumptions which will be used throughout the present work.

From now on, I set $2\pi\alpha'=1$, I ignored the overall factor
$T_p$ and an additive constant. The metric is Euclidean and is
``mostly plus''. That is,
$g_{\alpha\bar{\beta}}=\delta_{\alpha\bar{\beta}}$ and
$g_{\alpha\beta}=g_{\bar{\alpha}\bar{\beta}}=0$ Indices denoted by
$\mu,\nu,\ldots$ run from 1 to 2$p$, $i,j,\ldots$ run from 1 to
$2p$, $\alpha,\beta,\ldots$ run from 1 to $p$. There is no
distinction between upper and lower indices. All are lowered.
Repeated indices are summed over. Ordering of indices are also
disregarded. Anti-hermitian matrices for U($n$) generators (some
subcartan algebra) are chosen, etc. The fieldstrength is given by
\begin{equation}
    F_{{\alpha}{\bar{\beta}}}=\partial_{\alpha}A_{\bar{\beta}}-\partial_{\bar{\beta}}A_{\alpha}
\end{equation}
Instead of using real spatial coordinates $x^{\mu}$, complex
coordinates $z^{\alpha}$. I denote $F$ as
$F=F_{\alpha\bar{\beta}}$ unless otherwise indicated. Also,
$F_{\alpha\bar{\alpha}}=0$ considering that I am working on a flat
space. Furthermore,
$F_{\alpha\beta}=F_{\bar{\alpha}\bar{\beta}}=0$.

Using (\ref{sqrt}), the ABI action (\ref{abi}) becomes
\begin{eqnarray}
    &&\!\!\!\!\!\!\!\!\!\!\!\!\!\!\!\!\!\!\!\!\!\!
    \Gamma_{\mathrm{BI}}
    \!=\!-\!\!\int\!
        \sum^{6}_{d=2}d^{p+1}\sigma \,\,C^d F^d
\end{eqnarray}
and using (\ref{fn}) or (\ref{fnexp}) with traced invariants given
by (\ref{t}), the simplified ABI action relevant to
$D=4,6,8,10,12$ space-time dimensions in D$p$-branes is given by
\begin{eqnarray}
    &&\!\!\!\!\!\!\!\!\!\!\!\!\!\!\!\!\!\!\!\!\!\!
    \Gamma_{\mathrm{BI}}
    \!\!=\!-\!\!\int d^{p+1}\sigma\,\,%\,\,\,
    \left\{
        \frac{1}{2}T_2+\frac{1}{2}T_2F
    \right.
    %\nonumber\\&&
    %\!\!\!\!\!\!\!\!
    +\frac{1}{4}
    \left(
    \frac{1}{2}T^2_2
        +T_4
    \right)
    \nonumber\\&&
    \!\!\!\!\!\!\!\!\!\!\!\!\!\!\!\!
    +
    \frac{1}{4}
    \left(
    \frac{T^2_2}{2}
        +T_4
    \right)F
    %\nonumber\\&&
    %\!\!\!\!\!\!\!\!
    \left.
    +\frac{1}{2}
    \left(
    \frac{T^3_2}{24}+\frac{T^{11}_{24}}{4}+\frac{T_6}{3}
    \right)
    \right\}\label{simp}
\end{eqnarray}
after setting $C^k=1$. Hence, this is the most simplified ABI
action I can have with the assumption of no ordering, zero
derivatives in fieldstrength tensor and using the characteristic
equations I generated with the algorithm this work presented.

Deforming $F_{\alpha\bar{\alpha}}=0$ to
\begin{eqnarray}
    \left(1+\frac{1}{2}T_2
    +\frac{1}{4}
    \left(
    \frac{T^2_2}{2}
        +T_4
    \right)\right)F=0,
\end{eqnarray}
the Donaldson-Uhlenbeck-Yau condition acquires  an order $F^5$
correction. After some invariance restoration, the equations of
motion integrate to the ABI action (\ref{simp}) with
\begin{equation}
        \frac{1}{2}T_2
    +\frac{1}{4}
    \left(
    \frac{1}{2}T^2_2
        +T_4
    \right)
    +\frac{1}{2}
    \left(
    \frac{T^3_2}{24}+\frac{T^{11}_{24}}{4}+\frac{T_6}{3}
    \right).
\end{equation}
Modulo an undetermined overall multiplicative constant, this
exhibits the Born-Infeld action through order $F^6$, leading to a
uniquely fixed equation of motion. These results raise the
suspicion that the ABI action is the only deformation which allows
solutions on D$p$-branes at angles of the form
$F_{\alpha\bar{\alpha}}=F_{\alpha\beta}=F_{\bar{\alpha}\bar{\beta}}=0$.
 \cite{koe} This suspicion provides an important tool to probe the
 structure of the effective action that captures correctly the
 D-brane dynamics.
%this is how to do an unnumbered subsection
\subsection*{Acknowledgements}
I thank Emmanuel T. Rodulfo of De La Salle University for
suggesting me to apply the algorithm to an ABI action and for
providing me the Maple-routine based preprint of Robert Delbourgo
from University of Tasmania where the devised algorithm in the
present work was based to generate the same characteristic
equations as given in Eqns. (\ref{t})-(\ref{fnexp}) for
$d=2,3,4,5,6$.

\end{document}